\documentclass[aps,showpacs,twocolumn,preprintnumbers,nofootinbib] {revtex4-2}
\usepackage{graphicx,latexsym,amssymb,amsmath,color,multirow,mathrsfs,epsfig,bm}
\usepackage{changes}
\usepackage{hyperref}
\hypersetup{colorlinks=true,linkcolor=blue,filecolor=magenta,urlcolor=cyan}
\date{\today}

\begin{document}

\title{Pairing Strength and  Quadrupole-Soft Tin Isotopes}

    \author{Nguyen Le Anh}
	\email{anhnl@hcmue.edu.vn}
	\affiliation{Department of Physics, Ho Chi Minh City University of Education, 280 An Duong Vuong, Cho Quan Ward, Ho Chi Minh City, Vietnam}
    
   \author{Bui Minh Loc}
	\email{lmbui@sdsu.edu}
    \affiliation{San Diego State University,
    5500 Campanile Drive, San Diego, CA 92182}
    
    \author{Panagiota Papakonstantinou}
	\email{ppapakon@ibs.re.kr}
	\affiliation{Rare Isotope Science Project, Institute for Basic Science, Daejeon 34047, Korea}
    
    \author{Calvin W. Johnson}
	\email{cjohnson@sdsu.edu}
    \affiliation{San Diego State University,
    5500 Campanile Drive, San Diego, CA 92182}

    \author{Naftali Auerbach}
	\email{auerbach@tauex.tau.ac.il}
	\affiliation{School of Physics and Astronomy, Tel Aviv University, Tel Aviv 69978, Israel}

\begin{abstract}
\noindent
\textbf{Background:}
Understanding the experimental $B(E2)$ values for Sn isotopes around $^{110}$Sn has been a significant challenge in nuclear structure studies for over a decade. Both experimental data and many, though not all, calculations suggest a picture of the light Sn isotopes as being quadrupole-soft, that is, spherical, yet easy to deform.
\\
\textbf{Purpose:}
To investigate the delicate interplay of quadrupole deformation and pairing correlations in these nuclides. In particular, by using slightly enhanced pairing, we ask: can we generate spherical mean-field solutions that describe the data?
\\
\textbf{Method:}
First, we apply the standard spherical Skyrme HFBCS-QRPA calculation with default pairing parameters, allowing us to identify nuclides that are unstable  against quadrupole deformation 
among Sn isotopes. Next, we moderately enhance the pairing strength to reproduce the experimental binding energy in the deformation-unstable isotopes. \\
\textbf{Result:}
Within our choice of Skyrme parameters and use of density-independent pairing,
this moderate adjustment sufficiently stabilizes the HFBCS ground states against deformation, ensuring a successful QRPA calculation and, more importantly, leading to more realistic properties for the quadrupole $2^+$ states. 
\\
\textbf{Conclusion:}
Careful attention to the
sensitive interplay of pairing and shell effects
in deformation-soft nuclides can be crucial to their correct descriptions. This
sensitivity can be exploited to optimize the treatment of pairing in phenomenological approaches such as the present Skyrme-QRPA. 
\end{abstract}

\maketitle

\newpage
\section{Introduction}

The nuclear shape is largely governed by the interplay between shell structure and the quadrupole and pairing forces.
Quadrupole forces drive collective motion and deformation, while pairing forces and shell gaps stabilize the spherical shape by suppressing deformation. The balance between these forces determines whether a nucleus remains spherical, or undergoes a shape transition, driving structural evolution through the nuclear chart~\cite{BohrMottelson1998}.

Between doubly magic nuclei, which are spherical and hard to deform, and well-deformed nuclei, are `soft quadrupole deformed' nuclei.
Such nuclei are spherical, but the potential energy surface along the generator coordinate defined by the Bohr quadrupole deformation parameter $\beta_2$,  has a low curvature, meaning it is easy to deform them. This leads to quadrupole excitations low in energy. 
Neutron-deficient Sn isotopes are an excellent example of this class of nuclei

The extensive chain of experimentally accessible Sn isotopes provides an ideal testing ground for  this interplay. 
The known Sn isotopes include the entire neutron shell between the neutron magic numbers $50$ and $82$, while the proton magic number $50$ closure 
makes it easier, from a theoretical point of view, to isolate the effects of pairing and deformation coming from one nucleon species. 
However, the electric quadrupole transition strengths ($B(E2)$ values) observed in the lighter Sn isotopes as a function of neutron number $N$  pose a significant challenge to nuclear structure theories for decades--see Ref.~\cite{SicilianoPLB8062020} and references therein.
Specifically, the experimental transition strength between the first quadrupole state and the ground state in neutron-deficient isotopes are approximately constant, deviating
from the smooth parabolic behavior along the rest of the chain, suggesting a structure change, perhaps between a quadrupole and a pairing regime.

The shell model has been used to address 
the delicate counterbalance of quadrupole and pairing forces in Sn isotopes, for example in large-scale shell-model diagonalizations~\cite{SicilianoPLB8062020}, which suggested a rather strong quadrupole component  Sn isotopes regardless of pairing. 
This analysis upheld a closed $Z=50$ shell. 
Similarly, calculations using the Monte Carlo shell-model framework~\cite{TogashiPRL121} suggest dominance of quadrupole forces in the light isotopes and a weakening of the $Z=50$ shell. 
A more recent review of the topic, including the newest experimental data for $^{112,116,120}$Sn~\cite{BeuschleinPRC110}, pointed out that the presence or not of a dip in the measured $B(E2)$ at $^{116}$Sn could depend to some extent on the experimental techniques utilized. Based on the new data, it is concluded that the transition from the light isotopes' quadrupole-collective regime to the heavier isotopes' seniority regime is smoother than predicted by the calculations of Ref.~\cite{TogashiPRL121}. 
Finally, in hybrid \textit{ab initio} calculations (starting from a chiral effective field theory ($\chi$EFT) picture, carrying out Hartree-Fock-Bogolyubov calculations in a shell model framework, improved with many-body perturbation theory~\cite{scalesi2024EPJA}), the discrepancies observed between calculations and data 
are attributed to the weak pairing induced by the $\chi$EFT interaction, 
which in turn leads to the development of weak but finite quadrupole deformation, especially in the lighter Sn isotopes.

Many calculations of Sn isotopes utilize
the self-consistent mean-field approximation and the quasiparticle random-phase approximation (QRPA) on top of it. Constrained Hartree-Fock-Bogolyubov (HFB) calculations with a semi-realistic interaction 
have predicted
the neutron-deficient tin nuclei to be soft against  quadrupole deformation, with large $B(E2)$ values and near-flat potential energy curves \cite{OmuraPRC1082023}. 
According to that study, the result was triggered by a near-degeneracy of the $n0g_{7/2}$ and $n1d_{5/2}$ levels in these calculations and 
compromises the applicability of the spherical HFB plus QRPA approach.  
Note that the two levels are pseudo-spin partners and therefore their near-degeneracy could be expected, especially in a relativistic framework~\cite{LIANG20151}. 
On the other hand, the self-consistent relativistic HFB plus QRPA method using the NL3* functional \cite{Ansari2005PLB}  
has provided a qualitative agreement with data without considering ground-state deformation. 
As regards energy density functional theory for the ground state, we note that HFB calculations with the Gogny D1S interaction predict spherical shapes or very weak deformation in light Sn isotopes \cite{cea_potential}, while the deformed relativistic Hartree-Bogoliubov theory in continuum with the PC-PK1 energy density functional predicts spherical shapes for all even Sn isotopes \cite{Guo2024ADNDAT}. 
Quadrupole deformation parameters extracted from electric quadrupole moments and from transition probabilities to the first quadrupole state are also found equal to approximately 0.1, which is quite small \cite{stone2016ADNDAT,pritychenko2016ADNDAT}.

Most calculations support a spherical or near-spherical shape for all Sn isotopes between $A=100,132$, and therefore the applicability of QRPA in spherical symmetry. 
QRPA takes into account both pairing and a consistent treatment of the residual interaction. 
However, the QRPA has not yet been widely applied to this problem. 
In addition, the results of \cite{OmuraPRC1082023} in comparison with \cite{Ansari2005PLB}  raise the question of how robust the QRPA results are and what the role of the input effective interaction or functional is. 
Given the highly relevant role of the pairing interaction in the case of Sn isotopes and their quadrupole states, it seems worth looking into the role of pairing in the context of QRPA.

In this work, we will use the spherical Skyrme Hartree-Fock-Bardeen-Cooper-Schrieffer (HFBCS) plus QRPA method for the $2^+$ transitions in Sn isotopes and study their quadrupole softness. 
We augment the results and discussion with a self-consistent HFB-QRPA calculation using the Gogny D1S interaction~\cite{Berger1991}. 
We showed the connection between the instability of the spherical shape to the instability of the spherical QRPA result in Ref.~\cite{AnhPRC10924}. In the same spirit, here we are able to assess whether an Sn isotope is predicted to be soft or unstable against quadrupole deformation without the need for a deformed-QRPA application.

We encounter quadrupole instability in some isotopes depending on the functional input parameters. 
Interestingly, we find that a moderate adjustment of the pairing interaction can lead to stabilization of the spherical shape and physical results for the quadrupole states, an outcome that confirms the delicate interplay between correlations and pairing.

This paper is organized as follows. 
In Sec.~\ref{sec:QRPA} we present the theoretical method. 
In Sec.~\ref{sec:results}, we first confirm the diagnosis of quadrupole softness using representative Skyrme functionals, and next we show how a pairing adjustment can recover a more realistic picture. 
We conclude in Sec.~\ref{sec:conclusion}.

\section{Spherical QRPA calculation\label{sec:QRPA}}

The HF plus RPA and the HFBCS plus QRPA methods have been long in use \cite{rowe2010book, ring2004book} 
to calculate the response or susceptibility of nuclei to external perturbations. 
In Refs.~\cite{Colo2013, Colo2021}, publicly available codes for calculations in spherical symmetry based on Skyrme functionals 
are described. 
The methods are applied in a self-consistent way, {\em i.e.}, the same Skyrme functional 
which is used to describe the HF(BCS) ground state 
is also used to derive the (Q)RPA residual interaction between quasiparticle excitations. 
In addition, we will apply a self-consistent HFB-QRPA code for finite-range interactions~\cite{Hergert2011} such as Gogny.
Self-consistency along with large-enough model spaces is necessary to ensure the validity of known theorems in actual calculations~\cite{Thouless1960, Thouless1961}. 
Specifically, (Q)RPA restores the symmetries broken by the mean-field calculation for the ground state, for example, translational symmetry and rotational symmetry. 
In the case of calculations in spherical symmetry, application in deformed nuclei can lead to the ``collapse" result \cite{AbbasNPA3671981}, {\em i.e.}, the presence of imaginary solutions.  
Thouless's theorem tells us that the presence of an imaginary solution in, {\em e.g.}, the $2^+$ channel 
is equivalent to a prediction of a quadrupole-deformed ground state, even though the degree of deformation cannot be established. 
This property was exploited in our previous works on deformation softness for even-even nuclei throughout the nuclear chart~\cite{LocPRC1082023, AnhPRC10924}, and will also be utilized here. 

For studying genuinely deformed nuclei, appropriate implementations of (Q)RPA beyond spherical symmetry should be applied. On the other hand, if experimental data point to a spherical shape 
and a spherical QRPA calculation ``collapses", then the input to the (Q)RPA calculation, namely the functional parameterization, can be questioned. 
We hypothesize that this is the case for the neutron-deficient Sn isotopes: Although the measured properties of their $2^+_1$ state do not single them out in comparison to the rest of the isotopic chain, QRPA calculations with certain energy density functionals show them either very soft or deformed. 
Given that the functionals used here generally give a fair description of many other nuclei, especially magic ones, we are motivated to examine the role of pairing in this case. 

We employ the spherical Skyrme-HFBCS-QRPA calculation using the computer program {\tt hfbcs-qrpa} provided in Ref.~\cite{Colo2021}. We choose the SLy5 \cite{Chabanat1998}, SkM* \cite{BartelNPA3861986}, and SkP \cite{Dobaczewski1984_SkP} Skyrme parameter sets.

The computer program {\tt hfbcs-qrpa} provides as a preset default the density-independent contact pairing (or volume pairing) written as
\begin{equation}
    V^\delta(\bm r, \bm r') = V_0^i \delta(\bm r - \bm r'). \label{pairing}
\end{equation}
The preset values of $V_0^i$, in MeV$\cdot$fm$^3$, are $265.5$, $222.9$, and $213.7$ for SLy5, SkM*, and SkP, respectively. These values were obtained by fitting to the experimental neutron pairing gap in $^{120}$Sn ($\Delta_n = 1.245$ MeV) with a neutron pairing window fixed by taking into account only bound states above the neutron Fermi energy and to a maximum of 6 neutron s.p. levels. We refer to Eq.~\eqref{pairing} with preset values of $V_0$ as the \textit{default pairing}. In our previous work \cite{AnhPRC10924}, it was used to observe the nuclear deformation landscape, including quadrupole, octupole, and hexadecapole. Now, we use it as a starting point to study in more detail the quadrupole deformed softness in Sn isotopes.

The computer program {\tt hfbcs-qrpa} provides a more complex option for the density-dependent pairing, including surface and mixed pairing.
However, in view of the exploratory nature of the present work, we consider the simple prescription, Eq.~\eqref{pairing}, adequate for our purposes. The use of SkP is the same as SLy5 and SkM$^*$, i.e, HFBCS-QRPA with the separate phenomenological pairing force Eq.~\eqref{pairing}. It allows us to compare their results. The pairing strength $V_0^i$ is the only adjustable parameter in the study. We will modify its value, fit to binding energies rather than to a given pairing gap.

When addressing pairing, several issues must be considered. The first concerns the choice of the pairing interaction itself. One may employ a density-independent pairing force, as assumed in Eq.~\eqref{pairing}, or adopt more sophisticated forms such as density-dependent or surface-peaked pairing interactions. In some frameworks, pairing is not introduced as a separate force at all, but is incorporated directly into the effective interaction \cite{Dobaczewski1984_SkP}. Another issue is the treatment of pairing correlations. These can be handled within the BCS approximation, as in the present work, or through a fully self-consistent HFB approach \cite{Dobaczewski1984_SkP}. 
However, the simple prescription of Eq.~\eqref{pairing}, which relies on a single parameter, is found sufficient for illustrating the sensitivity of the results for the lowest quadrupole states of Sn to the pairing interaction. 

\section{Results and discussion\label{sec:results}}

\subsection{Diagnosing quadrupole softness and deformation}
First, we identify the soft or quadrupole-deformed Sn isotopes according to spherical QRPA calculations and using three Skyrme functionals as described above and in previous work~ \cite{AnhPRC10924}.
At this initial, diagnostic stage, we use the default pairing parameters found in the original distribution of the code {\tt hfbcs-qrpa}~\cite{Colo2021}.
We calculate the energy $E(2^+_1)$ and transition strength from the ground state $B(E2)$ for the first quadrupole state $2^+_1$. 
The $B(E2)$ values will be given in the Weisskopf unit (W.u.) so that we can compare the isotopes on an equal footing.
For the quadrupole transition, 1 W.u. $= 0.0594 A^{4/3}e^2\text{fm}^{4}$.

\begin{figure}
    \centering
    \includegraphics[width=0.5\textwidth]{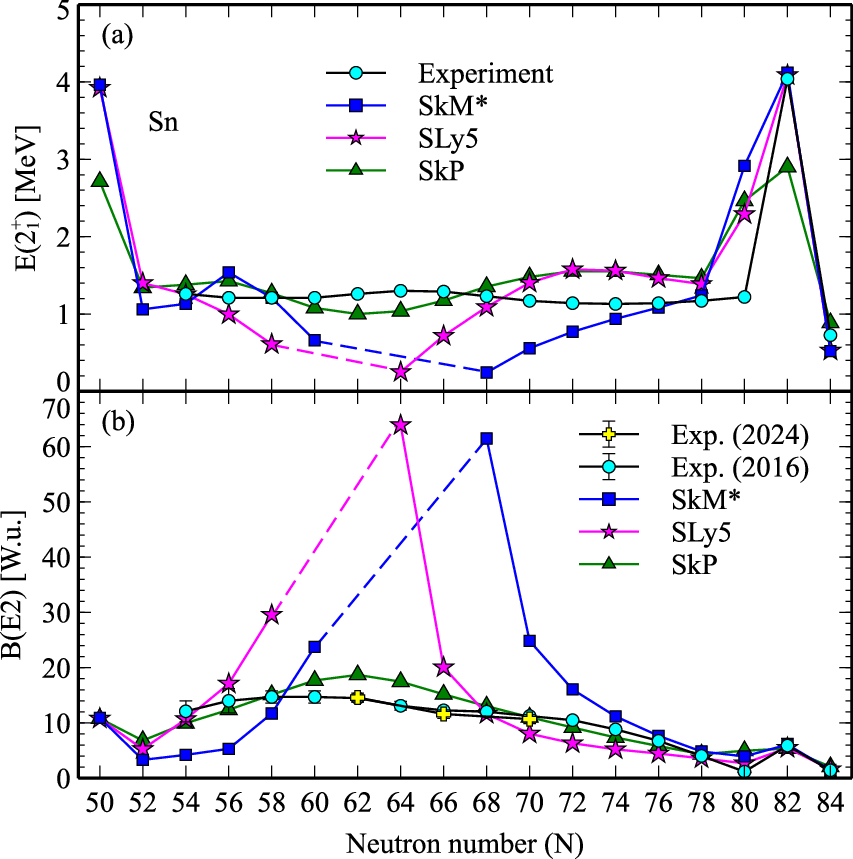}
    \caption{Diagnosing quadrupole-deformed softness in Sn isotopes using SLy5, SkM$^{*}$ and SkP with the default pairing parameters. 
    In the cases where the calculations lead to collapse (imaginary solutions), results are omitted and indicated by dashed lines. The experimental data were taken from Refs.~\cite{pritychenko2016ADNDAT, BeuschleinPRC110}.
    }
    \label{fig:Sn}
\end{figure}

The results are shown in Figure~\ref{fig:Sn}.
Using the SLy5 force, the $^{106-116}$Sn ($N = 56-66$ in Figure~\ref{fig:Sn}) are found soft or unstable against the quadrupole deformation. 
The indications are that the computed $E(2^+_1)$ values are found well below the flat experimental trend, indicating softness, or imaginary, indicating deformation. 
In the latter case, the values are not shown on the figure. 
The SkM* also predicts soft or deformed isotopes in this region, specifically $^{110-122}$Sn. 
The computed $B(E2)$ values fluctuate strongly, and as the excitation energy diminishes 
at $^{108}$Sn and $^{114}$Sn (SLy5) 
or at $^{110}$Sn and $^{118}$Sn (SkM*) 
they become too large, as expected from the conservation of the energy weighted sum rule. 

It is worth noting that the SkP parameter set did not give rise to deformed Sn isotopes.
SkP was developed with the explicit aim to describe accurately the pairing properties \cite{Dobaczewski1984_SkP}. 
The nucleon effective mass in SkP is set equal to the bare nucleon mass, and this leads to much different single-particle spectra, which in turn affects pairing. 
Its successful application in the present case points to the possible relevance of the pairing treatment in the description of the problematic Sn isotopes. 
We note that, unlike the shell-model calculation \cite{TogashiPRL121}, in the QRPA approach, the shape evolution in Sn isotopes is not linked with breaking the $Z = 50$ magic number. And unlike calculations of Ref.~\cite{OmuraPRC1082023}, 
all three functionals predict the $n0g_{7/2}$ and $n1d_{5/2}$ levels well separated from each other by roughly $2$~MeV. 
On the other hand, we find a near-degeneracy of the $3s_{1/2}$ and $2d_{3/2}$, which are also pseudo-spin partners in the $sdg$ shell. 
Thus, even in the present case, pseudo-spin symmetry may be playing a role. 

Next, we comment on predictions for the shape of Sn isotopes from some other HFBCS and HFB calculations and how our present results compare to those. 
We note that even when the same functional is used, it is not possible to compare our numerical results with other calculations on an equal footing, owing to the differences in handling the pairing channel and the fact that we rely on the QRPA calculation instead of the HFBCS or HFB to diagnose softness.
Previous calculations using the SLy5 and SkM* functionals, but with different pairing strength $V_0$ and a three-dimensional HFB code, which is different from the spherical HFBCS code used here \cite{Colo2021}, obtain spherical Sn isotopes \cite{TerasakiPRC782008}. It is not clear whether QRPA calculations built on those reference states would yield softness or instabilities. 
To examine this aspect of the problem further, we have performed fully self-consistent spherical HFB-QRPA calculations with the Gogny D1S interaction~\cite{Berger1991}. 
We have used a highly self-consistent computational code such that spurious state energies are of the order of tens of keV~\cite{Hergert2011}. 
We compared the isotopes predicted to be deformed by this HFB-QRPA calculation and those predicted to be deformed by the deformed-HFB calculations reported in \cite{cea_potential}. 
We found instabilities between and including $^{104}$Sn and $^{120}$Sn. 
On the other hand, on a grid of deformation parameters $\beta$ with a spacing of $0.05$, the tabulated values in \cite{cea_potential} report deformed ground states for $^{110,112,120}$Sn only. It is not clear if all of the other isotopes are predicted to be exactly or approximately spherical with $|\beta| < 0.05$. In any case, the comparisons confirm an extreme sensitivity to computational details.

The delicate interplay between pairing and deformation suggested by the above results motivates a closer look at pairing in an attempt to obtain stable spherical QRPA results for Sn isotopes with other Skyrme forces. Here, we present an example with the SLy5 force. 
Qualitatively similar results can be obtained with other forces. 

\subsection{Treating the softness with pairing-reinforced QRPA}

\begin{figure}
    \centering
    \includegraphics[width=0.5\textwidth]{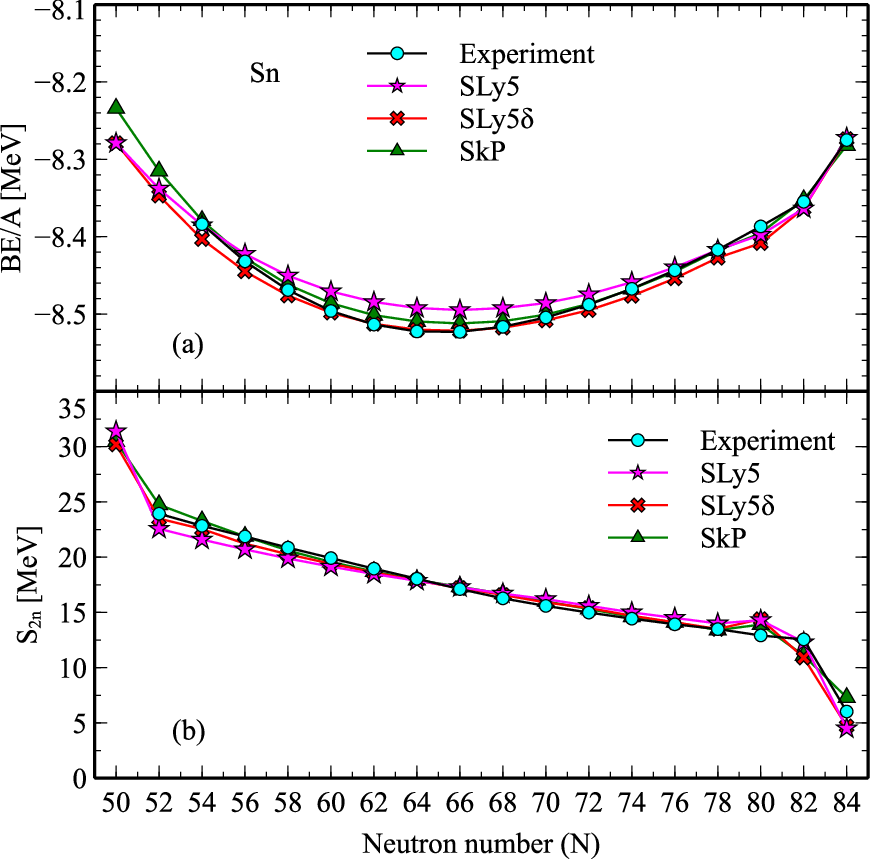}
    \caption{Results for the binding energy per particle (a) and the two-neutron separation energies (b) with different Skyrme forces, including the pairing-reinforced SLy5 force (SLy5$\delta$).} 
    \label{fig:SnSLy5}
\end{figure}

Next, we increase the pairing strength to reproduce exactly the experimental binding energy per nucleon of $^{112}$Sn using the SLy5 interaction. $^{112}$Sn lies within the instability region in Figure~\ref{fig:Sn}.
An increase of the pairing strength parameter from $265.5$ to $334.5$~MeV$\cdot$fm$^3$ changes the calculated binding energy per particle of $^{112}$Sn from $-8.484$~MeV to $-8.513$~MeV. 
The SkP functional yields $-8.502$~MeV. 
Figure~\ref{fig:SnSLy5} shows the binding energies per particle BE/A and two-neutron separation energies $S_{2n}$. 
Generally, the trends are smooth, while with the new pairing, an improvement in the $S_{2n}$ is observed in comparison to the default pairing. 

\begin{figure}
    \centering
      \includegraphics[width=0.5\textwidth]{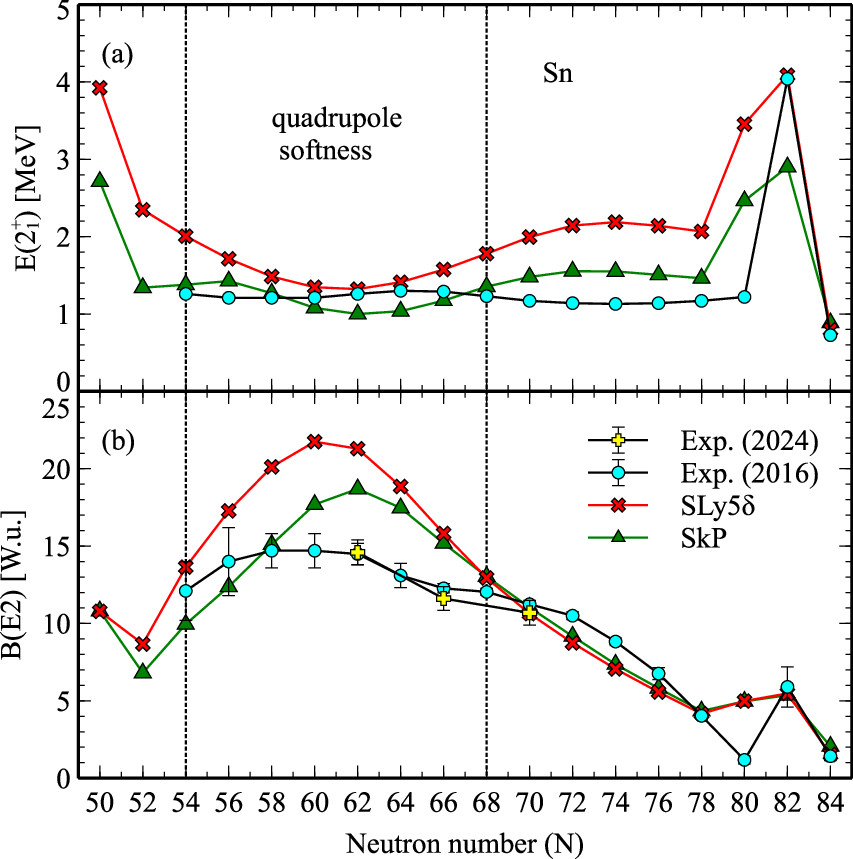}
    \caption{Results for the Sn isotopes ($N = 58-66$) with the pairing-reinforced QRPA using SLy5 force (SLy5$\delta$) compared with data and with the results using the original SkP force.}
    \label{fig:finalizedResult}
\end{figure}

Figure~\ref{fig:finalizedResult} shows the $2^+_1$ transition in comparison with the data and with the results using the SkP interaction, which we have left unchanged. 
The pairing reinforcement has brought the results almost on a par with the SkP results, and there are no more instabilities. 
A better agreement with data would, in fact, be achieved with a smaller adjustment to the pairing parameter.

It is worth noting that the HFB method using the NL3* functional \cite{Ansari2005PLB}
requires a stronger pairing force to describe the binding energies in the light Sn isotopes but not in the heavier ones, and therefore, the author advocated a pairing adjustment that could cure this discrepancy. This is consistent with the present results. 

\begin{figure}
    \centering
    \includegraphics[width=0.5\textwidth]{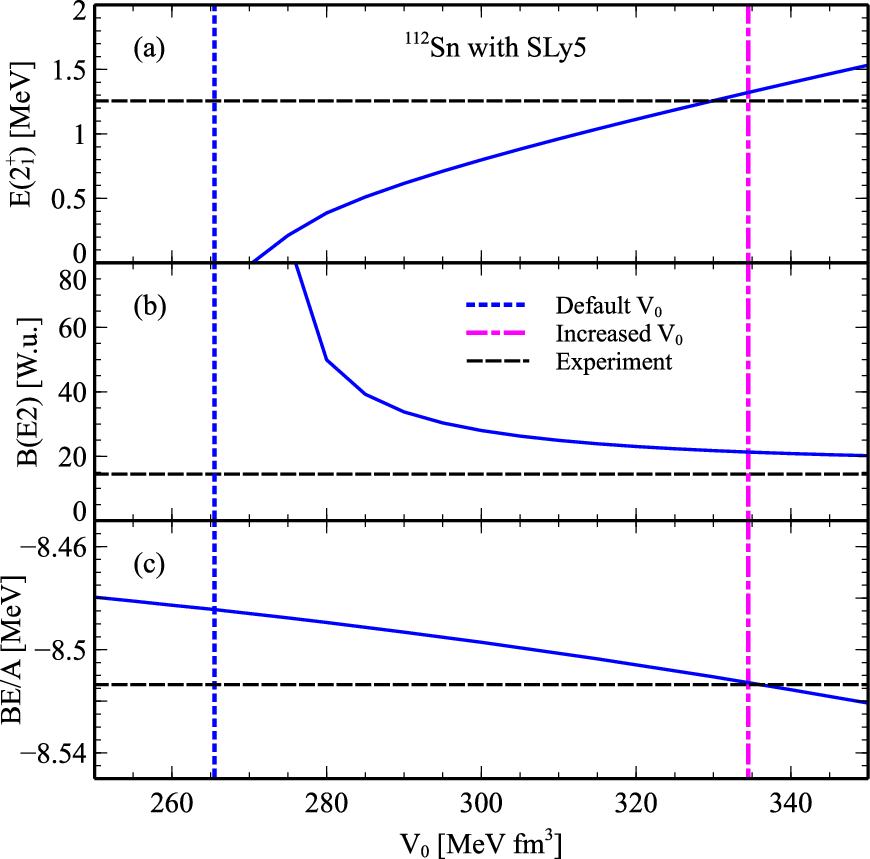}
    \caption{The energy $E(2^+_1)$ (a), the transition strength $B(E2)$ (b), and the binding energy $BE/A$ (c) in the case of $^{112}$Sn using the SLy5 force as functions of the pairing strength $V_0$. The default value and the increased value of $V_0$ are marked by vertical lines. The horizontal dashed lines indicate the experimental values from Ref.~\cite{stone2016ADNDAT}.} 
    \label{fig:V0_112Sn}
\end{figure}

\begin{figure}
\includegraphics[width=0.5\textwidth]{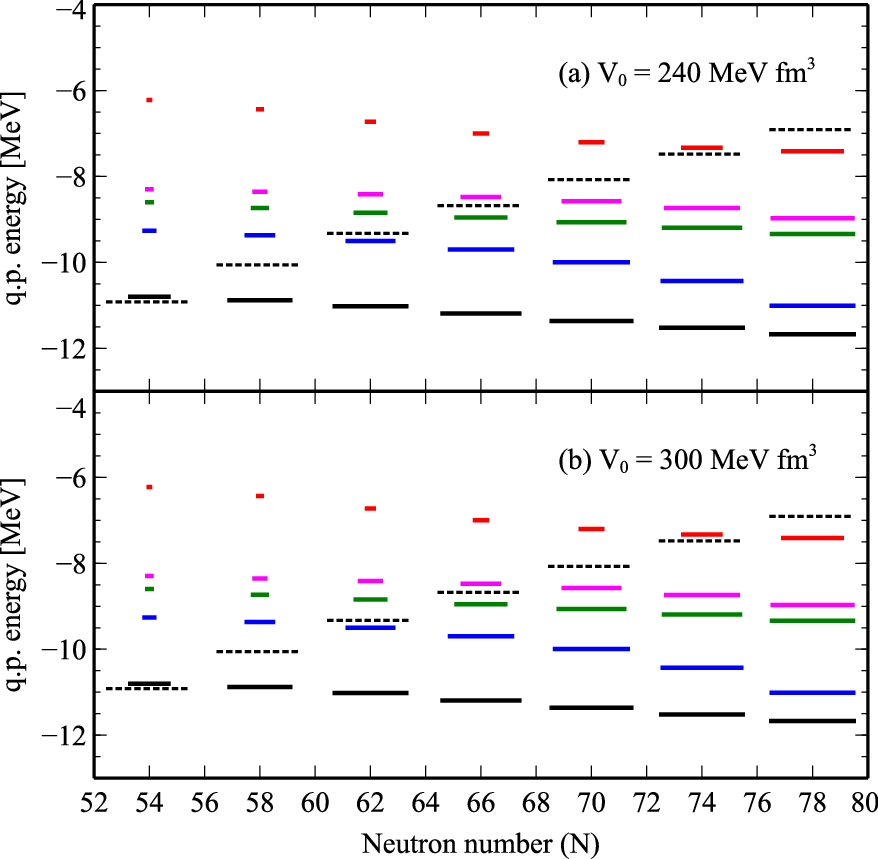}
\caption{Evolution of single-neutron energies and occupation probabilities along the Sn isotopic chain when using the SLy5 force and different values of the pairing strength $V_0$. From lowest to highest energy are $2d_{5/2}$, $1g_{7/2}$, $3s_{1/2}$, $2d_{3/2}$, and $1h_{11/2}$. For clarity, results are shown for every two even isotopes. The length of each bar corresponds to the occupation probability of the state. A dotted line represents the Fermi energy.} \label{fig:spstates}
\end{figure}

Let us now take a look at how the weak change of one parameter can have such a drastic effect in certain nuclei in the context of self-consistent (Q)RPA calculations. 
Figure \ref{fig:V0_112Sn} shows the energy and transition strength of the first $2^+$ state as well as binding energy per particle (c) in $^{112}$Sn calculated with the SLy5 force and varying pairing interaction strength $V_0$.
The binding energy varies very smoothly and moderately. The energy of the first $2^+$ state varies smoothly for large enough values of $V_0$, but as $V_0$ is lowered, it reaches zero and becomes imaginary. At that critical value of $V_0$, the solution collapses to a deformed state. 
The transition strength $B(E2)$ rises dramatically as the energy approaches zero, as is expected from the conservation of the energy weighted sum rule in QRPA, given that, as we have verified, the giant resonance region is very weakly affected. 
The general smoothness of the trends shows that it is the binary character of {\em collapse} or {\em no collapse} at the near-critical conditions encountered in these nuclei that leads to the apparent sensitivity of the results shown in Figs. \ref{fig:Sn} and \ref{fig:finalizedResult}. 
On the other hand, the HF-BCS single-particle spectrum itself is not much affected by the pairing strength as shown in Figure~\ref{fig:spstates}. 
Therefore, the sensitivity lies rather in the competition of the QRPA residual interaction and the pairing interaction. 
As previously mentioned, we find that the pseudo-spin partners $n0g_{7/2}$ and $n1d_{5/2}$ are well separated, but $n2d_{3/2}$ and $n3s_{1/2}$ are not, and the near-degeneracy could also drive instability depending on the residual interactions.  

The question arises as to whether such a pairing adjustment in the Sn isotopic chain can spoil the results for other nuclei and their agreement with data. 
Small but non-negligible changes in the binding energies are, of course, to be expected in open-shell nuclei, as 
Figure~\ref{fig:SnSLy5} already suggests. 
The effect will not be problematic in a protocol that takes into account open-shell nuclei in the parameter fit and, at the same time, ensures a good description of softness-prone Tin isotopes. 

As regards the properties of $2^{+}_1$, an inspection of results in the Ca isotopes, which represent a lighter mass region, has revealed only a moderate effect. If softness is indeed driven by closely placed single-particle levels, along with a mismatched pairing interaction, then it should manifest in heavier mass regions rather than in lighter ones. For example, for Pb isotopes beyond $N=126$, the $spgi$ shell becomes active and contains three such partners, starting with $n0i_{11/2}, n1g_{9/2}$. We found that $^{210}$Pb and mid-shell isotopes, for which at present no data exist, are predicted to be unstable towards quadrupole deformation by SLy5 with the default pairing. It is not possible to make a direct attribution of the instabilities to shell effects and pairing, but our results indicate that the issue of exaggerated quadrupole softness may not be restricted to the Tin isotopic chain. 

\section{Conclusion\label{sec:conclusion}}

Because of the sensitive interplay between the quadrupole interaction and pairing, some
nuclear models can predict soft quadrupole transitions in neutron-deficient Sn isotopes.
In the present work, we have used the self-consistent QRPA method, which takes into account both the pairing interaction and the full Skyrme interaction, to diagnose quadrupole softness and instability in Sn isotopes and examine their origins. 
Our results confirm the critical role of the pairing and suggest that the near-degeneracy of pseudo-spin partners in the neutron $sdg$ shell might play a role.
The observed sensitivity to the handling of pairing correlations could be exploited in future optimizations of phenomenological approaches such as Skyrme models.

\section*{Acknowledgements}
P.P was supported by the Institute for Basic Science
(IBS) through the NRF (2013M7A1A1075764). This material is based upon work supported by the U.S. Department of Energy, under Award Number DE-NA0004075.

\bibliographystyle{apsrev4-2}
\bibliography{refs.bib}

\end{document}